\documentclass[aps,prb,amsmath,twocolumn,amssymb,floatfix,superSCriptaddress,footinbib]{revtex4-1}
\pdfoutput=1
\usepackage[dvips]{graphics}
\usepackage{bm}
\usepackage{epsfig}
\usepackage{enumerate}
\usepackage{subfigure}
\usepackage{color}
\usepackage{amsmath,amssymb,amsthm}
\usepackage{sidecap}
\usepackage{graphicx}
\usepackage{hyperref}
\hypersetup{
    pdfnewwindow=true,      
    colorlinks=true,       
    linkcolor=blue,          
    citecolor=magenta,        
    filecolor=blue,      
    urlcolor=blue        
}

\newcommand{\ie}{{\it i.e., }}

\newcommand{\bea}{\begin{eqnarray}}
\newcommand{\eea}{\end{eqnarray}}
\newcommand{\beq}{\begin{equation}}  
\newcommand{\eeq}{\end{equation}}

\begin{document} 
\title{Signature of odd-frequency equal-spin triplet pairing in the Josephson current on the surface of Weyl nodal loop semimetals} 

\author{Paramita Dutta}
\email{paramita.dutta@physics.uu.se}
\affiliation{Department of Physics and Astronomy, Uppsala University, Box 516, S-751 20 Uppsala, Sweden}
\author{Annica M. Black-Schaffer}
\email{annica.black-schaffer@physics.uu.se}
\affiliation{Department of Physics and Astronomy, Uppsala University, Box 516, S-751 20 Uppsala, Sweden}

\begin{abstract}
We theoretically predict proximity-induced odd-frequency (odd-$\omega$) pairing on the surface of a Weyl nodal loop semimetal, characterized by a nodal loop Fermi surface and drumhead-like surface states (DSSs), attached to conventional spin-singlet $s$-wave superconducting leads. Due to the complete spin polarization of the DSS, only odd-$\omega$ equal-spin triplet pairing is present, and we show that it gives rise to a finite Josephson current. Placing an additional ferromagnet in the junction can also generate odd-$\omega$ mixed-spin triplet pairing, but the pairing and current are not affected if the magnetization is orthogonal to the DSS spin polarization, which further confirms the equal-spin structure of the pairing.
\end{abstract}

\maketitle

\section{Introduction}
Weyl nodal loop semimetals (WNLSs) are recently discovered topological semimetals, where the valence and conduction bands cross each other along a closed one-dimensional ($1$D) loop carrying a $\pi$ Berry flux~\cite{Barderson,FuTopo,Burkov,RMNweylsup,KimRingNode,matsuura2013protected,Kawazoe,parhizgarWNLS,Chen_weyl}. The band topology results in drumhead-like surface states (DSSs) whose boundary is defined by the projection of the Fermi nodal loop onto the two-dimensional ($2$D) surface Brillouin zone (BZ)~\cite{Oded,nandkishore2016weyl}. There already exist several proposals for WNLSs, including SrIrO$_3$ based on crystal symmetry analysis~\cite{chen2015topological}, PbTaSe$_2$ based on topology analysis~\cite{PbTaSe}, TITaSe$_2$ and HgCr$_2$Se$_4$ based on first principles~\cite{TITaSe,HgCrSe}, and tight-binding calculations~\cite{Oded}. There are also reports of experimental characterization, for PbTaSe$_2$ using angle-resolved photoemission spectroscopy~\cite{PbTaSe} and Ca$_3$P$_2$ using x-ray diffraction~\cite{xie2015new}. 

The topology of WNLSs makes them a very interesting candidate for unconventional superconductivity. WNLSs have already been shown to allow for fully gapped chiral three-dimensional (3D) bulk superconductivity\,\cite{nandkishore2016weyl}, while $p+ip$ chiral surface superconductivity has been proposed to be present in the DSS if the bulk has $p$-wave pairing\,\cite{RMNweylsup}. On the other hand, the DSS is fully spiin polarized for all WNLSs breaking time-reversal symmetry, making it completely immune to proximity-induced spin-singlet pairing from a conventional $s$-wave superconductor (SC). However evidence exists evidence in the literature showing that instead odd-frequency (odd-$\omega$) spin-triplet $s$-wave pairing can appear in simple ferromagnetic materials \cite{Rmp,MultilayerFS}. This motivates us to investigate if the DSS also induces odd-$\omega$ pairing on the surface of a WNLS in proximity to a conventional SC.

Odd-$\omega$ superconductivity refers to when two electrons in a Cooper pair are odd in the relative time coordinate, or equivalently frequency \cite{Rmp,linder2017odd}. It was first predicted by Berezinskii~\cite{BerezinskiiHe3} in the context of $^3$He and later introduced for superconductivity~\cite{kirkpatrick1991tr,BalatskyNewclass,Schrieffer1994} and also the Kondo lattice~\cite{coleman1994odd}. Research on systems hosting this unusual pairing has flourished during the last few decades~\cite{linder2017odd}. One major reason behind this is that the fermonic nature of the Cooper pairs allows for unusual pairing symmetries, such as odd-$\omega$ spin-triplet $s$-wave or spin-singlet $p$-wave pairing.

Odd-$\omega$ superconductivity has mainly been considered for hybrid structures\,\cite{PDfs,PDnsn,PDTI} like ferromagnet (FM)-SC~\cite{linder2015superconducting, di2015signature, linder2015strong, Rmp,tanakaFS, MultilayerFS, Matsumoto2013}, normal metal (NM)-SC~\cite{TanakaNSjunction,qtLinder,CayaooddwRashba}, and 
topological insulator-SC\,\cite{AnnicaTI,BjornTI, Kuzmanovski_oddw, CayaooddwTI}, but has also been considered in multi-band SCs with finite inter-band hybridization~\cite{AnnicaMulti,AnnicaMulti2}, and conventional Josephson junctions,\,\cite{balatsky2018odd} as well as for SCs subjected to time-dependent drives\,\cite{TriolaDriven,TriolaMDriven}. Odd-$\omega$ behavior is also present for systems with Majorana fermions\,\cite{OddMajorana}. Experimentally, this exotic pairing state can be captured through various experimental phenomena like the Meissner effect\,\cite{TanakaMeissner1,BalatskyMeissner,TanakaMeissner2,EsrigMeissner}, Josephson current\,\cite{TanakaJosephson}, Majorana scanning tunneling microscopy,\cite{MajoranaSTM}, Kerr effect\,\cite{komendova2017odd,TriolaUPt3}, and thermopower measurement\,\cite{thermo}. 

\vskip -0.018cm
In this work we use a Josephson junction setup on the surface of a WNLS to explore proximity-induced superconductivity, especially odd-$\omega$ pairing. Odd-$\omega$ pairing was very recently discussed in regard to WNLSs in Ref.\,\onlinecite{parhizgarWNLS}, but with leads attached across the bulk of the WNLS, thus measuring the bulk effects. We instead us	e a much simpler setup on a single surface and show explicitly how only odd-$\omega$ equal-spin triplet pairing is present on the surface, due to the complete spin polarization of the DSS. We further prove that the odd-$\omega$ pair amplitude is directly measurable by generating a finite Josephson current. Moreover, the flat-band dispersion of the DSS is also detectable in the Josephson current, as it forces part of the current to flow into sub-surface layers and reduces the overall magnitude. By adding a FM region to the junction we find additional mixed-spin triplet pairing, but only when the FM magnetization opposes the spin polarization of the DSS. The finite proximity effect and Josephson current on WNLS surfaces can therefore be entirely attributed to odd-$\omega$ pairing, with its equal-spin structure verified by anisotropic behavior with an external magnetic field. 

\section{Model and density of states}
The Hamiltonian for a minimal model of a $3$D WNLS breaking time-reversal symmetry is \cite{Oded,matsuura2013protected,Kawazoe}
\bea
\bm{H}_{\text{w}} (\bm{k})&=&\bm{\sigma_x} (6 - \alpha_1 - 2 \cos k_x - 2 \cos k_y - 2 \cos k_z)  \nonumber \\
&&+  2 \alpha_2 \bm{\sigma_y} \sin k_z - \mu_{\text{w}}
\label{Ham_w}
\eea
where $k_i$ is the wave vector along the $i$-th ($x$, $y$ or $z$) axis and the Pauli matrices, $\bm{\sigma_i}$ act in spin space. 
The first two terms give rise to a Fermi nodal loop with its shape set by $\alpha_1$ and $\alpha_2$ and protected by $\mathcal{TI}$-symmetry, a product of time-reversal ($\mathcal{T}$) and inversion ($\mathcal{I}$) symmetry, 
and mirror symmetry with respect to $k_z$$\rightarrow$$-k_z$\,\cite{RMNweylsup}. For the finite chemical potential $\mu_{\text{w}}$ the Fermi surface is instead a small torus-like surface surrounding the nodal loop. We here use $\alpha_1$$=$$\alpha_2$$=$$1$ and $\mu_{\text{w}}$$=$$1$ (in units of the nearest neighbor hopping integral), but the appearance of odd-$\omega$ pairing is not restricted to these values.

For surfaces perpendicular to the $z$ axis, the Fermi nodal loop projects to form a large topologically protected DSS. To capture the DSS we therefore consider a finite cubic lattice system along the $x$ and $z$-axes, but keep the periodicity along  the $y$ axis to arrive at $\bm{H}_{\text{w}}(k_y)$, (see Appendix~\ref{DisW}). This leads to the nearest-neighbor hopping integrals $-\bm{\sigma}_x$ and $-\bm{\sigma_x}$$-i$$\alpha_2$$\bm{\sigma}_y$, along  the $x$ and $z$-axes, respectively. We use a slab $L_z$$=$$21$ layers thick along the $z$ axis with $L_x$$=$$16$ sites along the $x$ axis for each layer. 

To form a Josephson junction on the surface we attach two conventional spin-singlet $s$-wave SCs to the top layer 
\begin{figure}[htb]
\centering
\includegraphics[scale=0.51]{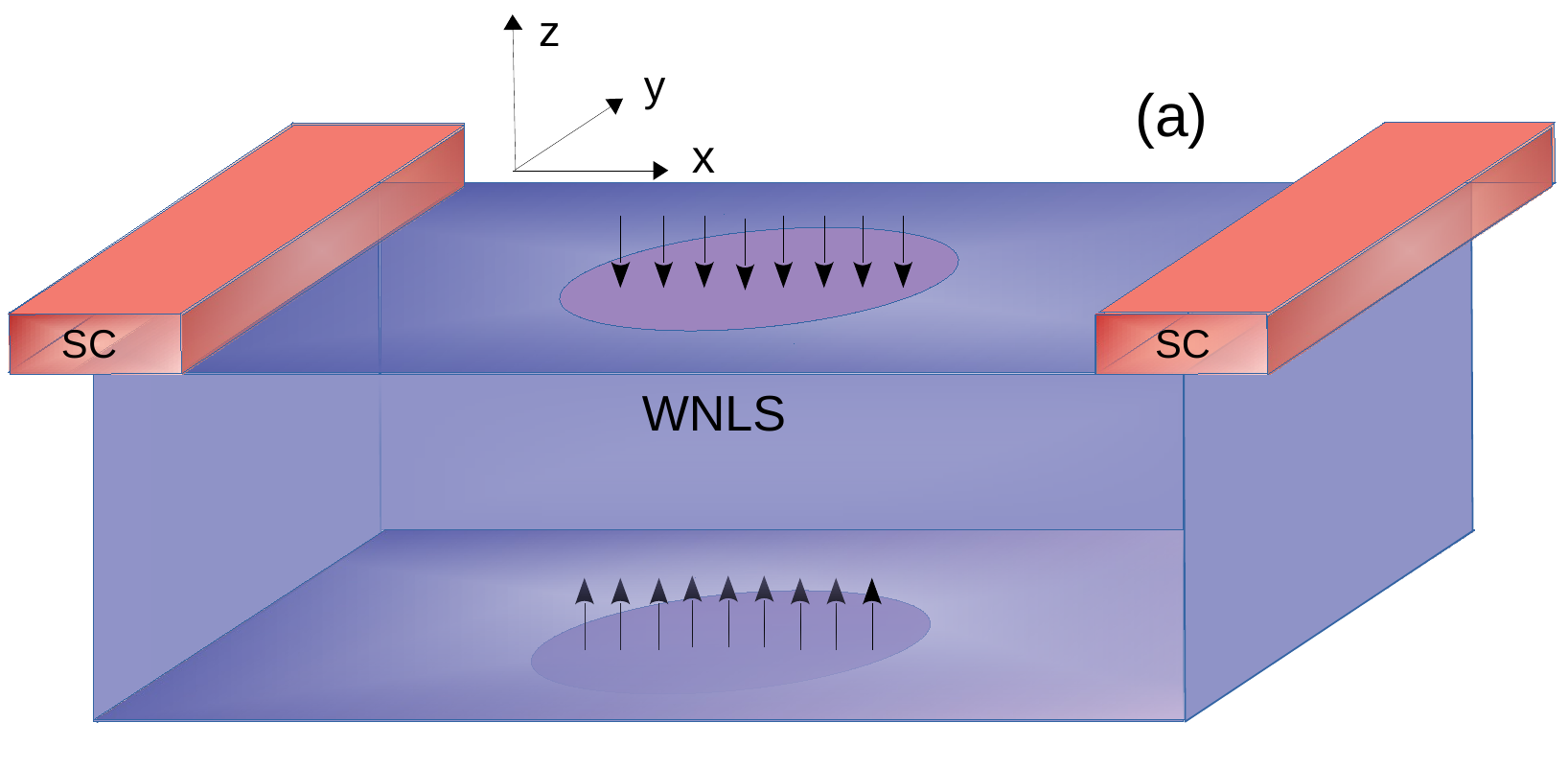}
\includegraphics[scale=0.65]{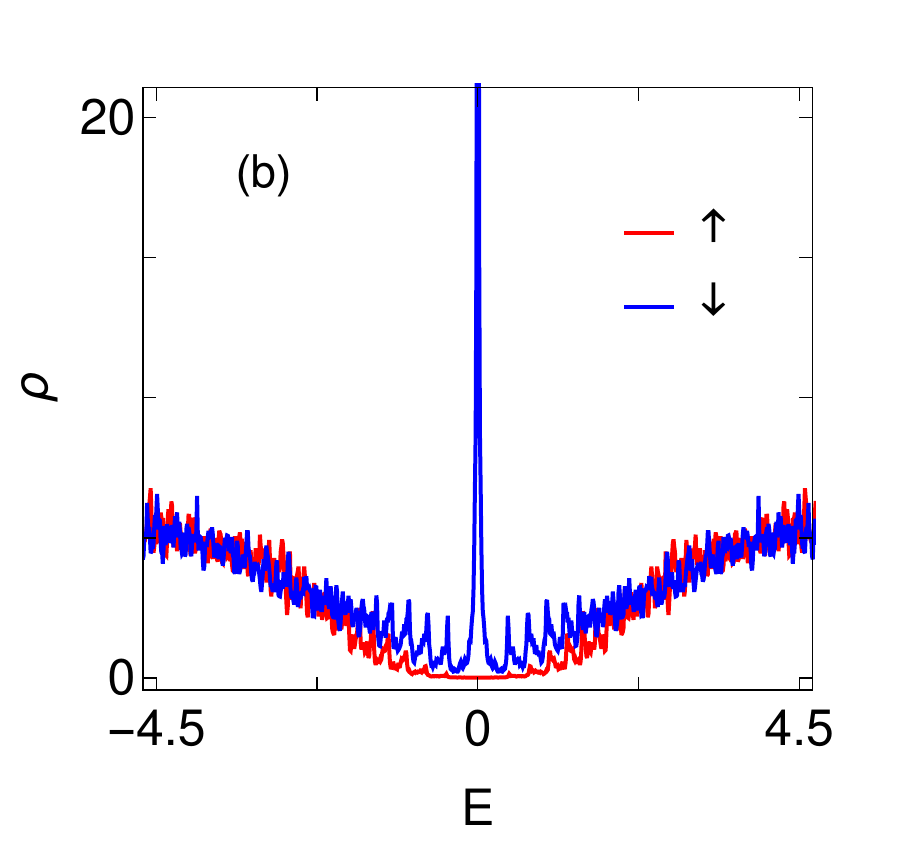}
\caption{(a) Schematic figure of a WNLS with two SC leads attached to the surface. The spin polarization of the DSSs (circular regions) of the top and bottom layers is shown by arrows. (b) SLDOS at the middle site of the top layer of the WNLS as a function of energy $E$.}
\label{model}
\end{figure}
($n_z$$=$$1$) of the WNLS, at the $n_x$$=$$1$ and $n_x$$=$$L_x$ sites [see Fig.\,\ref{model}(a)]. Each lead is described by a finite number of lattice sites along the $x$ axis keeping the periodicity along the $y$ axis to mimic %
bulk SCs. For the SC and tunneling Hamiltonians, we refer to Appendix~\ref{anoGn}. The left and right leads are characterized by the order parameter $\Delta_s$$=$$0.01$ and phase factors, $\phi_L$ and $\phi_R$, respectively. This results in a superconducting coherence length $\xi\,(=\hbar v_{Fx}/\Delta$)\,$\sim 64 a$ based on the $x$ axis bulk Fermi velocity. The tunneling amplitude between the WNLS and each lead is denoted by $t_{\text{w-sc}}$. We also add a FM island inside the junction (sites $n_x$$=$$3$ to $n_x$$=$$14$) by adding the on-site term $m_n$$(\bm{P} \cdot {\bm{\sigma}}$) to $\bm{H_{\text{w}}}$. 
The polarization vector of the FM is $\bm{P}$ ($\sin \theta_F \cos \phi_F,\sin \theta_F \sin \phi_F,\cos \theta_F$), while $m_n$ sets the magnitude. 

We calculate the spin-resolved local density of states (SLDOS) at different lattice sites on the surface of the WNLS {using the retarded Green's function of the bare WNLS to define the SLDOS as 
\bea
\rho_{r,\sigma}(E)=-\frac{1}{\pi} \sum \limits_{k_y} \text{Im}[\bm{\mathcal{G}}^R_{r \sigma r \sigma}(E,k_y)],
\label{rho}
\eea
where $\bm{\mathcal{G}}^R_{r \sigma r \sigma}(E,k_y)$ is the element of the whole retarded Green's function given by
\bea
\bm{\mathcal{G}}^R(E,k_y)=[(E+i \delta) \bm{I}-\bm{H}_{\text w}(k_y)]^{-1}.
\eea
Here $r\,(=\{x,z\})$ is the lattice site index, and $E$ is the energy. 
To verify the existence of the DSS we show in Fig.\,\ref{model}(b) the calculated SLDOS polarized along the $z$ axis for the top layer of a finite WNLS, which has a very sharp peak at $E$$=$$0$ with complete spin-down polarization (for a complementary analytical calculation, see Appendix~\ref{SS}). Away from $E$$=$$0$, the SLDOS becomes almost linear and has comparable spin polarizations, signaling bulk contributions. On the bottom layer we find a similar surface state but with up-spin polarization as presented in Appendix~\ref{SS}.

\section{Odd-frequency pairing}
We next attach the SC leads and study the proximity-induced Cooper pairs on the WNLS surface. To capture the superconducting pair amplitude we consider the whole Josephson junction, the WNLS plus two SC leads, and calculate the anomalous Green's function\,\cite{BalatskyMeissner} (for details, see Appendix~\ref{anoGn}). To analyze the nature and the amplitude of the induced pairings, we plot in Fig.\,\ref{anoF} the real and imaginary parts of the pair amplitude $F$ as a function of frequency $\omega$ on the surface of the WNLS. All possible spin configurations of the pairing function, spin-singlet ($\uparrow\downarrow$$-$$\downarrow\uparrow$), mixed-spin triplet ($\uparrow\downarrow$$+$$\downarrow\uparrow$), and equal-spin triplet ($\uparrow\uparrow, \downarrow\downarrow$), are shown. We 
\begin{figure*}[!thpb]
\centering
\includegraphics[scale=0.5]{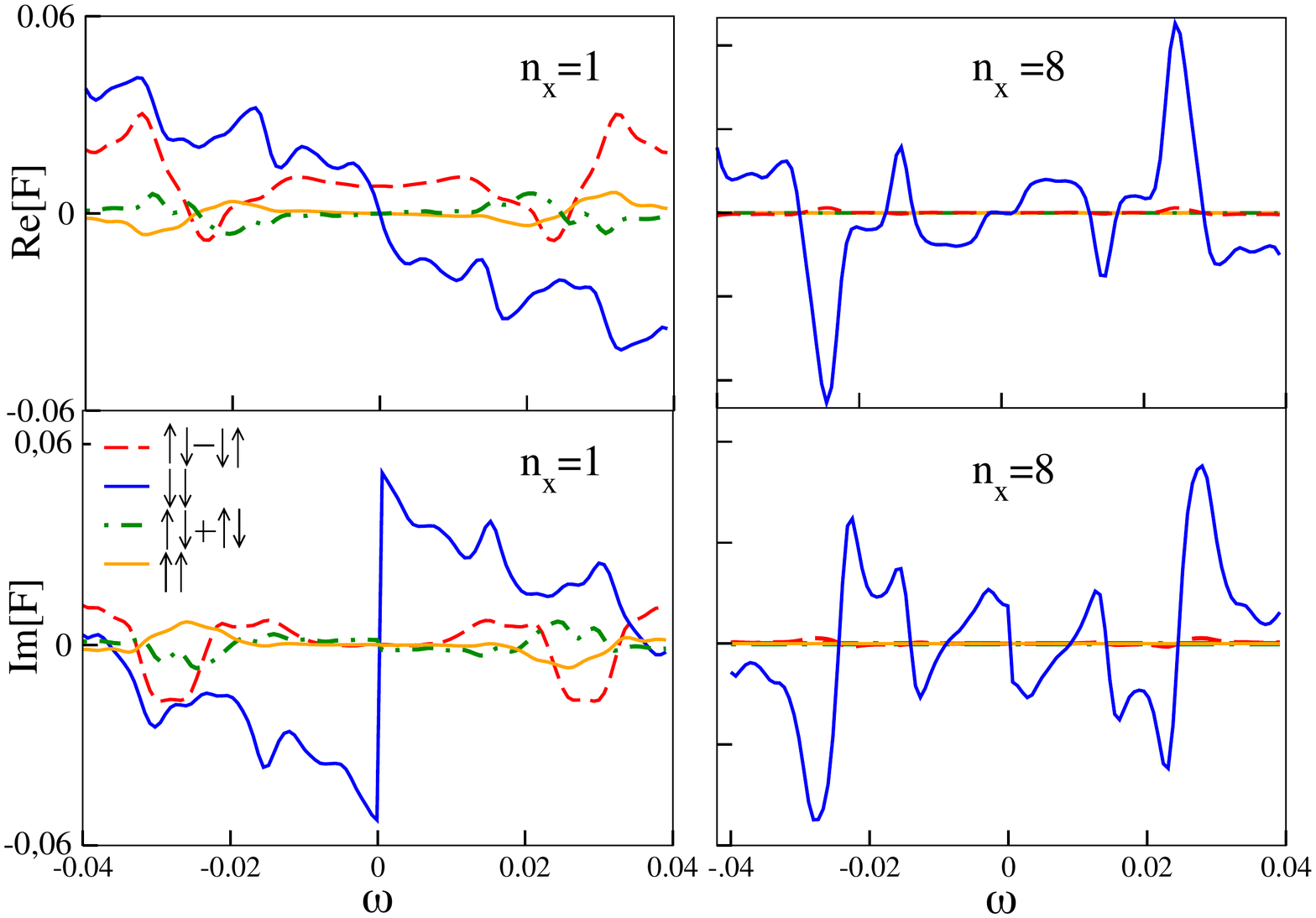}
\caption{Real (top) and imaginary (bottom) parts of $F$ as a function of frequency $\omega$ for two sites in the top layer ($n_z$$=$$1$) using $t_{\text{w-sc}}$$=$$0.5$.}
\label{anoF}
\end{figure*}
here consider only on-site $s$-wave pairing as we find the $p$-wave amplitudes to be negligible on the WNLS surface. Moreover, $s$-wave symmetry is very generally considered to be the stablest pairing in the presence of any disorder\,\cite{pwave-tanaka}. Due to the symmetrical positions of the SC leads on the surface, $F$ is symmetric with respect to the middle site. We therefore show only the results for the first ($n_x$$=$$1$) and the middle sites ($n_x$$=$$8$). We observe that spin-singlet pairing is present where the spin-singlet SC leads are located. However, this amplitude decays extremely quickly into the junction. Instead, the only pair amplitude present well inside the junction is the $\downarrow\downarrow$ spin-triplet pairing, all other amplitudes are negligible. 

The results in Fig.~\ref{anoF} have to be understood in the context of the complete spin-down polarization of the DSS. This spin polarization strongly opposes spin-singlet pairing, although exactly at the SC leads spin-singlet pairing still survives to some extent, as also observed in SC-FM junctions \cite{eschrig2008triplet}. 
Instead, the DSS supports only $\downarrow\downarrow$ pairing as tunneling of down spin is heavily favored compared to up spin from the perspective of energy cost. Following Fermi-Dirac statistics, any spin-triplet $s$-wave state necessarily has to have an odd-$\omega$ dependence. This is confirmed in Fig.~\ref{anoF}, where all spin-triplet pair amplitudes, both the real and imaginary parts, are odd functions of frequency. Similarly, we find only odd-$\omega$ equal-spin triplet pairing in all sub-surface layers, a consequence of the  spin polarization of the DSS. 
It is the topologically protected DSS and, particularly, its spin polarization that give rise to the equal-spin triplet odd-$\omega$ pairing. 

Odd-$\omega$ spin-triplet pairing is well established in SC-FM heterostructures\,\cite{Rmp,MultilayerFS,di2015signature}, including SC-half-metal structures \cite{tanakaHM,eschrig2008triplet}. As such, the appearance of odd-$\omega$ pairing on the WNLS surface is not surprising. However, for the long-range 
\begin{figure*}[htb]
\centering
\includegraphics[scale=0.5]{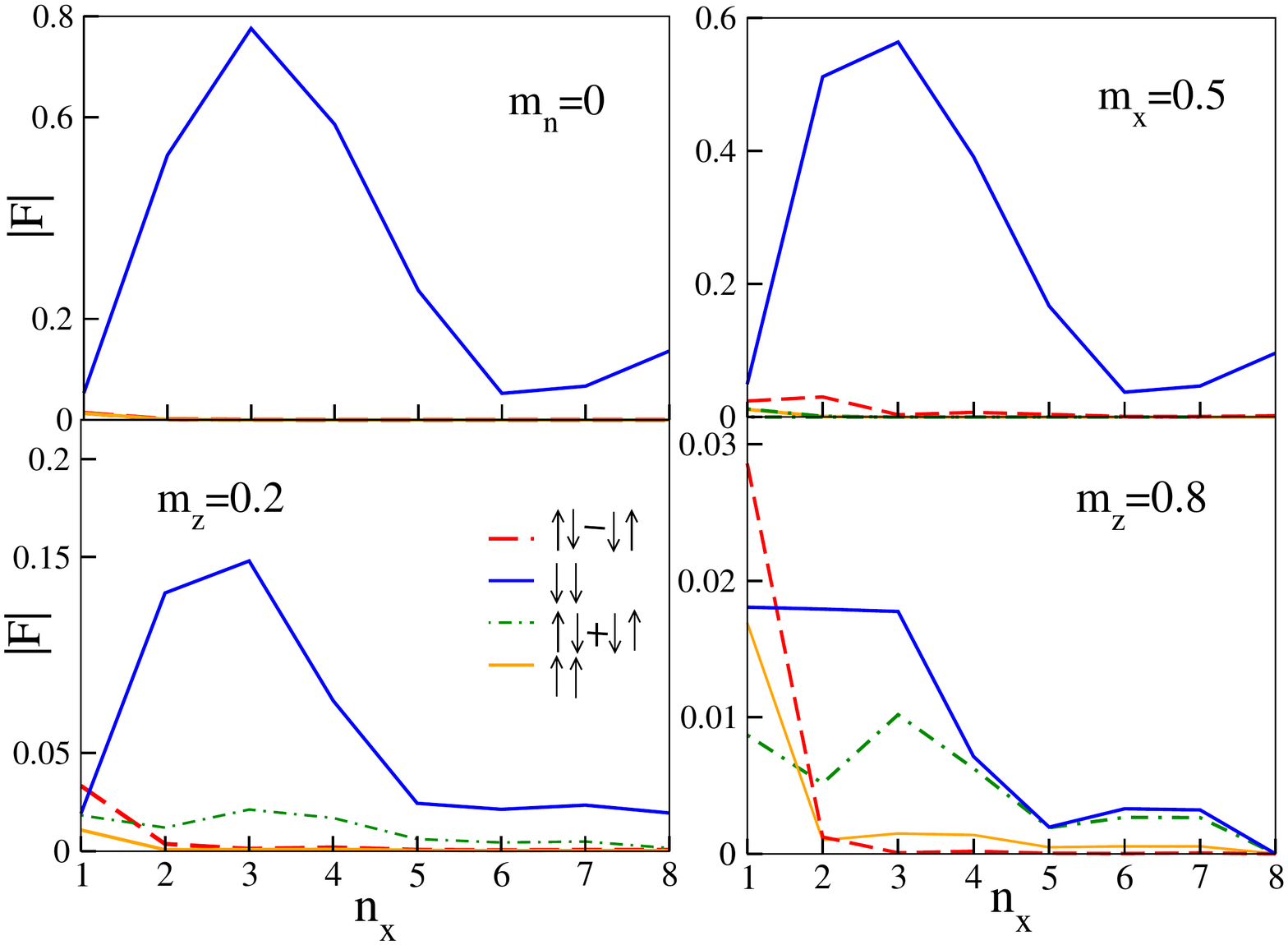}
\caption{Pair amplitude $|F|$ at $\omega$$=$$0.025$ as a function of $n_x$ in the top layer for different FM $\bm{P}$. Other parameters are the same as in Fig.\,\ref{anoF}.}
\label{Fnx}
\end{figure*}
equal-spin triplet pairing to appear in any of these previously studied structures, two magnetization directions have to be present: one magnetization direction rotates the spin-singlet state to a mixed-spin triplet state, while the second direction, not parallel to the first, generates equal-spin pairing. Alternatively, spin-orbit coupling can substitute for one of the magnetic fields, where for example an FM-SC interface with spin-orbit coupling has been shown to be sufficient to generate equal-spin triplet pairing \cite{eschrig2008triplet,Reeg,Bobkov}. In contrast, for a Josephson junction on the surface of a WNLS, the DSS is only polarized in only a single direction, here along the $-z$-direction, but still equal-spin triplet pairing completely dominates. We can attribute this to the existence of spin-orbit coupling present in the bulk of the WNLS, which is in proximity to the DSS. The WNLS surface Josephson junction therefore provide a unique system for generating long-range odd-$\omega$ pairing, as it does not require engineering a spin-active interface, nor any application of external magnetic fields, and still the pairing is exclusively of odd-$\omega$ equal-spin nature. This is one of the main results of this work. We also note that this result is distinctive from the recently found odd-$\omega$ pairing in the bulk of the WNLS, as spin-orbit coupling is automatically present inside the Josephson junction \cite{parhizgarWNLS}. As $\bm{H}_{\text{w}}(\bm{k})$ is symmetric with respect to the $x$ and $y$-axes, we conclude that odd-$\omega$ pairing is dominating irrespective of the orientation of the Josephson junction. 

\section{Ferromagnetic junction}
Having established the existence of only odd-$\omega$ equal-spin pairing in the DSS, we next add a FM island on the top layer of WNLS to further study the pairing structure. In Fig.\,\ref{Fnx}, we show the variation of the absolute values of $F$ as a function of the position in the junction $n_x$ in both the absence and presence of a FM. With no FM (i.e.\,$m_n$$=$$0$), the odd-$\omega$ $\downarrow\downarrow$ spin-triplet pairing completely dominates over all other pairings by an order of magnitude or more. As we move away from the SC leads towards the middle of the junction, the pair amplitude first increases due to the large spin polarization of the DSS, but as the distance from the SC lead increases the pair amplitude eventually decays. 

For a FM with a $\bm{P}$ vector along the $x$ axis this behavior is unaffected. We here show the result for $m_x$$=$$0.5$, but this is true for all values of $m_x$ and also $m_y$. However, the pairing changes if we set $\bm{P}$ in the $+z$-direction, i.e.\,opposite to the DSS spin polarization. The amplitude of the $\downarrow\downarrow$ pairing is then reduced. Instead, spin-singlet and also $\uparrow\uparrow$ spin-triplet pairing grows close to the SC leads. But, most importantly, the amplitude of the mixed-spin triplet pairing strongly increases and eventually becomes comparable to that of the down equal-spin pairing throughout the junction when we increase $m_z$.

We can understand the behavior of the WNLS with an added FM from the traditional FM Josephson junction, where the spin-singlet pairing of the SC is transformed to mixed-spin triplet pairing in the FM region \cite{Rmp,MultilayerFS,di2015signature}. But in the WNLS we see this effect only for a FM with $\bm{P}$ along $+z$, because then the FM counteracts the spin polarization of the DSS such that the down equal-spin pairing triplet is reduced, leaving space for other pairing symmetries. The reduced overall spin polarization also causes the spin-singlet pairing to increase close to the SCs. On the other hand, for $\bm{P}$ along the $x$ or $y$-directions, the FM acts on both spins equally, and thus the DSS can unperturbedly continue to generate the down equal-spin triplet pairing. As a consequence, the transformation from even-frequency (even-$\omega$) spin-singlet to odd-$\omega$ spin-triplet pairings depends on the value of $m_n$ as well as the direction of $\bm{P}$. Most notably, the mixed-spin triplet state appears only if a magnetization is present to counteract the intrinsic DSS spin polarization,
otherwise only equal-spin pairing is present, which is a very different behavior from traditional FM Josephson junctions. For the results in Fig.\,\ref{Fnx} we choose a particular $\omega$ where the amplitude of odd-$\omega$ is reasonably high, but the results do not change for other $\omega$, see Appendix~\ref{FM-w}. 

As seen in Fig.~\ref{Fnx}, there is only a appreciable change in the pair amplitude for $\bm{P}$ vectors with a component along the $z$ direction. Now, in Fig.~\ref{coupling} we show how these pair amplitudes behave in the 
weak and strong coupling limits in the presence of the FM island. We choose two values, $m_z=0.2$ and $0.8$, of the FM polarization similar to those in Fig.\,\ref{Fnx}. As seen, there is no qualitative difference in the nature of the pairing amplitudes between these limits: for both weak and strong coupling the mixed-spin triplet pairing increases with 
\begin{figure*}[thb]
\centering
\includegraphics[scale=0.53]{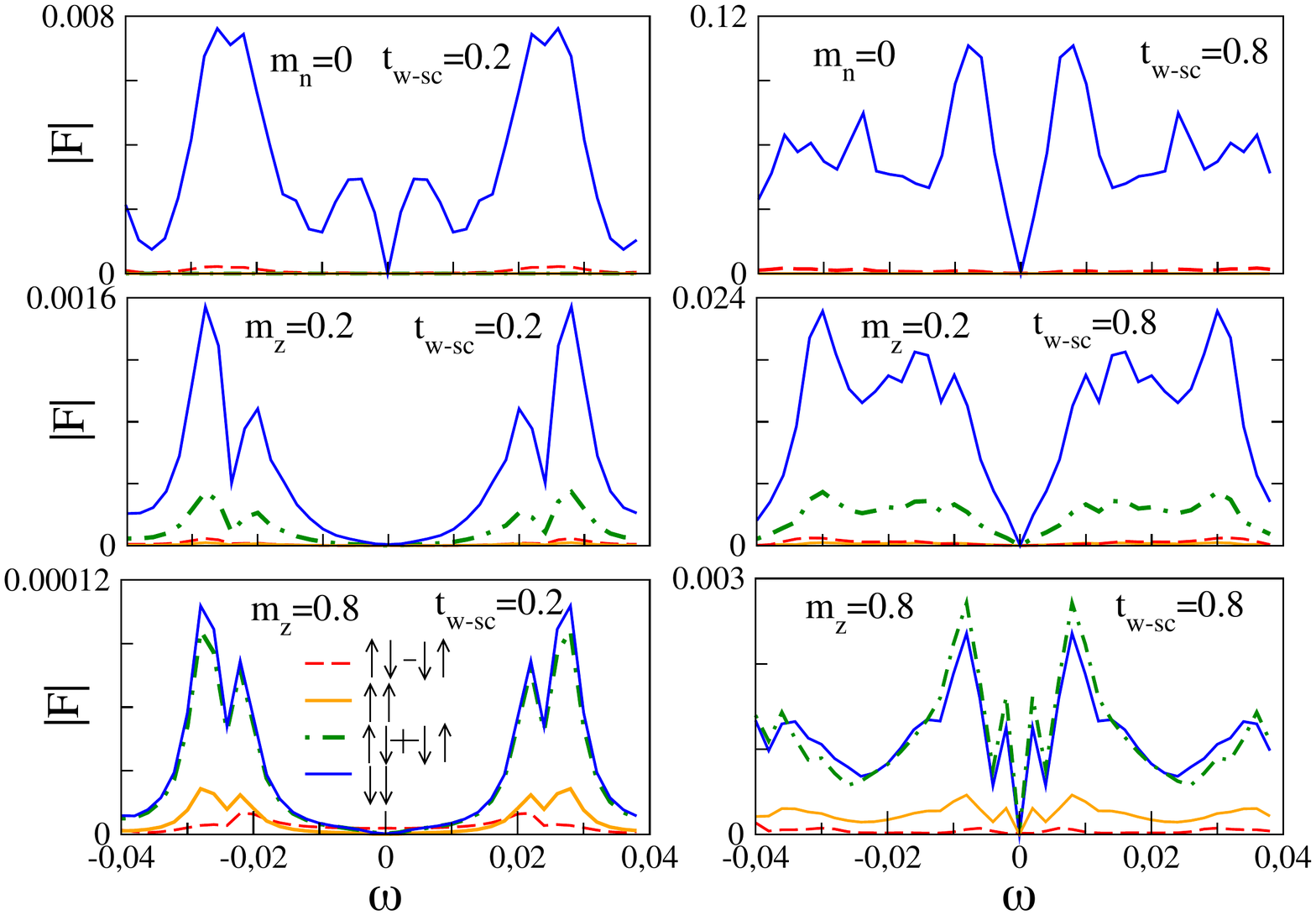}
\caption{Magnitude of pair amplitude $|F|$ as a function of frequency $\omega$ at the middle site ($n_x$$=$$8$) of the top layer ($n_z$$=$$1$) of  the WNLS in the absence ($m_n=0$) and in the presence of FM set along the $z$ axis. The left and right columns correspond to the weak and strong coupling regime, $t_{\text{w-sc}}=0.2$ and $t_{\text{w-sc}}=0.8$, respectively. Other parameter values are the same as in Fig.\,\ref{anoF} .}
\label{coupling}
\end{figure*}
$m_z$. In the strong coupling limit we see overall larger pair amplitudes compared to the no FM case, which explains the slightly larger currents in this regime.

\section{Josephson current}
To be able to measure the effect of odd-$\omega$ pairing, we calculate the Josephson current $J$, which appears 
for finite phase differences between the two leads, using the charge continuity equation\,\cite{BlackSchaffer2008,Covaci,Kristofer}, see Appendix\,\ref{calJ} for more details. We choose the phase difference $\Delta \phi$$=$$\pi/2$ as the current is maximum for this value (see Appendix~\ref{I-phi}) for the profile of $J$ vs. $\Delta \phi$. We express $J$ in units of $t_w \hbar/ea$ and plot it as a function of $t_{\text{w-sc}}$ in Fig.\,\ref{curr}. With increasing $t_{\text{w-sc}}$, $J$ grows due to the enhancement of the induced SC pairing, but ultimately saturates. 

In the absence of a FM, the current can be carried by only the odd-$\omega$ down spin-triplet pairing, as that is the only pairing present well inside the junction. From the previous studies of both conventional FM Josephson junctions (with two magnetization directions) \cite{di2015signature} and bulk WNLS junctions \cite{parhizgarWNLS}, we know that odd-$\omega$ equal-spin triplet pairing can carry a significant current. Based on the large DOS of the 
DSS we could also naively expect a very large current for a WNLS surface Josephson junction. To investigate this, we compare with a normal metal (NM) junction using the same SC leads, see Appendix~\ref{nm} for the NM Hamiltonian. From Fig.\,\ref{curr} we see that the current is actually suppressed in the WNLS junction compared to the NM junction, despite the much larger normal-state DOS for the WNLS. We attribute this effect to the (almost) flat-band nature of the DSS. With the carrier velocity being proportional to the band dispersion, $\partial E/\partial k_x$, the flat-band nature of the DSS limits the Josephson current, despite the large number of available carriers. 

To illustrate how the flat dispersion of the DSS suppresses the current, we divide in the inset of Fig.\,\ref{curr} we divide the total current into the contributions from each layer of the WNLS slab along the $z$ direction. Normally, we expect the current to be strongly concentrated to the first layers closest to the surface, as proximity-induced pairing quickly decays when moving away from the SC leads. However, in the middle of the junction ($n_x$$=$$8$) the current 
\begin{figure}[thb]
\centering
\includegraphics[scale=0.75]{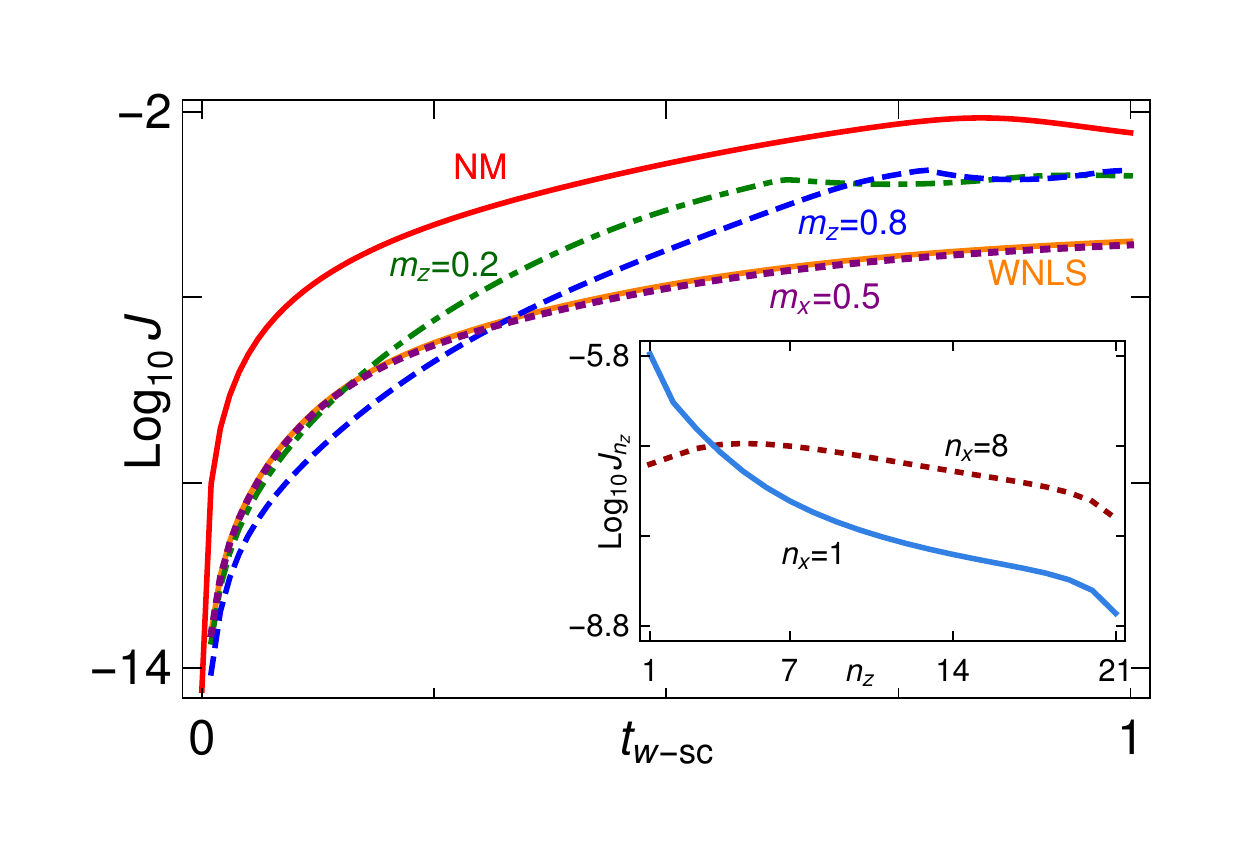}
\caption{Maximum Josephson current $\text{log}_{10}J$ as a function of of tunneling coupling $t_{\text{w-sc}}$ for a WNLS Josephson junction with and without a FM island, and for a NM junction. Inset: $\text{log}_{10}J_{n_z}$ (current in layer $n_z$) vs.~distance from the surface $n_z$ at the SC lead (solid) and in the middle of the junction (dashed). }
\label{curr}
\end{figure}
stays approximately constant for many sub-surface layers, even slightly increasing. This demonstrates that the DSS does not conduct the majority of the current, but the current instead flows through sub-surface layers despite this resulting in a much longer path between the two contacts. Thus, while the DSS enables a pure odd-$\omega$ 
Josephson current in all sub-surface layers, it also suppresses its magnitude in the surface layer, and in the junction as a whole. The only exception is for the site just below the SC ($n_x$$=$$1$), where the presence of the SC mitigates the effects of the DSS to give a more conventional behavior. This constitutes the other important main
result in this work: measuring a finite Josephson current proves the presence of odd-$\omega$ pairing, as no current would be present if only even-$\omega$ pairing were present. In addition, a majority of current in sub-surface layers, with an overall reduced magnitude, verifies the existence of extremely low-velocity carriers in the DSS.

Now, we add the FM island and show the total current for three different $\bm{P}$ in Fig.\,\ref{curr}. Keeping $\bm{P}$ along the $x$ or $y$ axis does not change the current at all. This is expected as this does not cause any fundamental change in the pair amplitudes. It is only for FM islands with a $\bm{P}$ component along the $z$ axis 
\begin{figure}[!thpb]
\centering
\includegraphics[scale=0.6]{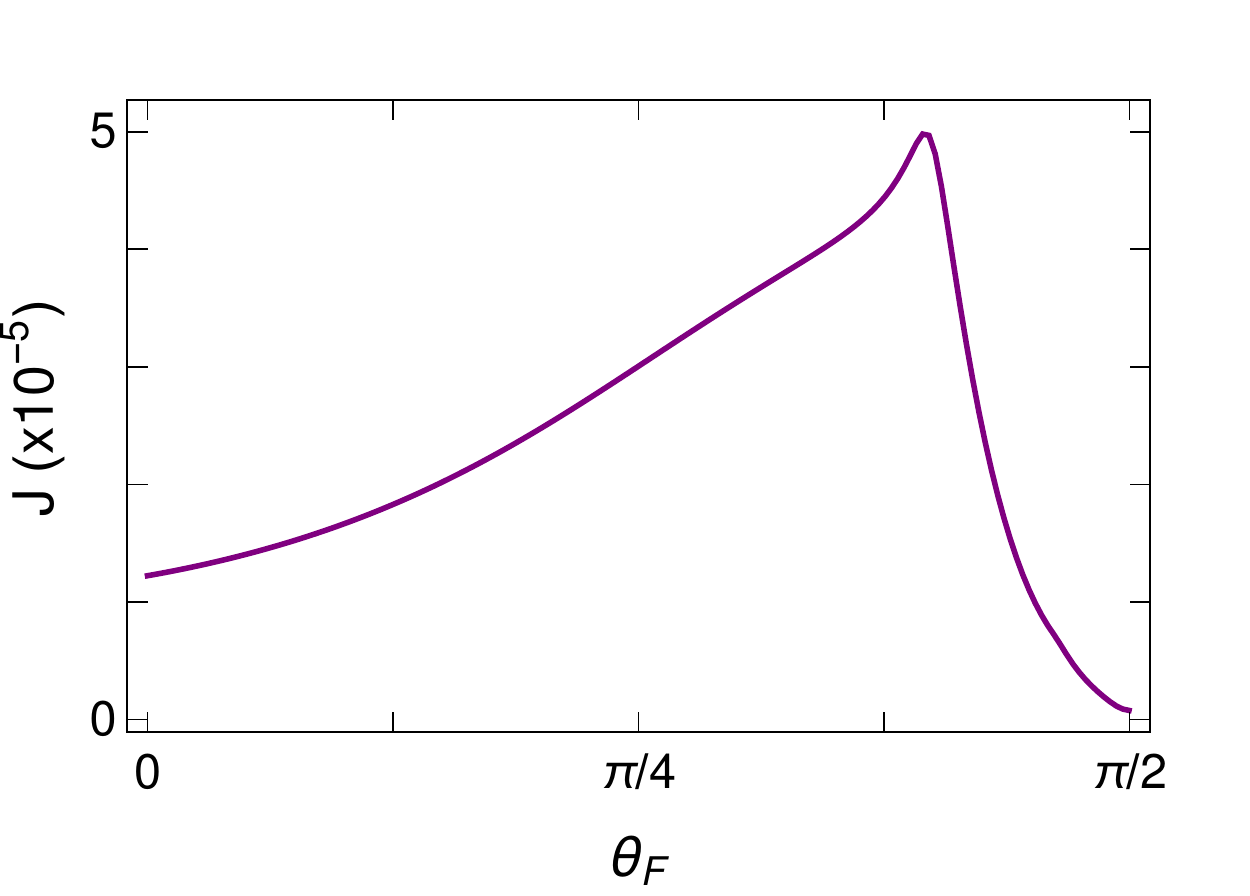}
\caption{Current $J$ as a function of the polarization polar angle $\theta_F$ for a FM island with $m_n$$=$$0.5$ and $t_{\text{w-sc}}=0.5$. Other parameter values are the same as in Fig.\,\ref{anoF}}
\label{Jxz}
\end{figure}
that the current is modified. For $\bm{P}$ along the $+z$-direction we see a moderate change in the current, also for large values of $m_z$ where mixed-spin triplet pairing is comparable to the equal-spin triplet pairing. 

Finally, we report how the current changes as the magnetization is tuned from the $z$ direction to the $x$-direction, given by varying $\theta_F$ from zero to $\pi/2$. In Fig.\,\ref{Jxz} we plot $J$ as a function of $\theta_F$ for $m_n$$=$$0.5$ and $t_{\text{w-sc}}$$=$$0.5$. We notice that with the increase of $\theta_F$, $J$ initially increases, and as $\bm{P}$ gets closer to the $xy$ plane the current again decreases and settles to the value found in the absence of a FM in Fig.~\ref{curr}. This non-monotonic behavior cannot simply be explained by the qualitative behavior of the pair amplitudes alone, but is the result of a combination of currents carried by the equal-spin and mixed-spin pairing. When $\theta_F$ increases, so does the $\downarrow\downarrow$ pair amplitude rather dramatically, while at the same time the mixed-spin amplitude is slowly decaying. At $\theta_F$$=$$\pi/2$, there are only $\downarrow\downarrow$ pairs, and the current is thus slightly reduced. Most importantly, the detailed numerical account of the behavior of the current for FM islands with a finite $\bm{P}_z$ component provides further evidence that the current is carried by equal-spin triplet pairs.

\section{Summary}
In this work we showed that the superconducting proximity-effect on the surface of WNLS is significant and consists entirely of odd-$\omega$ equal-spin triplet pairing, due to the complete spin polarization of the topologically protected DSS. The odd-$\omega$ pairing is directly measured by the presence of a finite Josephson current. The flat-band dispersion of the DSS is also visible by forcing significant current to flow into sub-surface layers and causing a reduction of the total Josephson current. In fact, the sub-surface current directly indicates the flat band nature of the surface states. Placing a FM island in the Josephson junction additionally generates mixed-spin triplet pairing when the FM magnetization opposes that of the DSS polarization. However, for FM magnetization perpendicular to the DSS spin polarization, the pairing and Josephson current remain unaffected, further corroborating the equal-spin triplet nature of the pairing in WNLS Josephson junctions. We note in particular that it is the topology of the WNLS that leads to the appearance of a pure odd-$\omega$ equal-spin triplet Josephson effect, without any further interface engineering needed. The equal-spin triplet supercurrent can be utilized in superconducting spintronics devices \cite{eschrig2011spin}, with the additional possibility of tuning the spin structure by a FM.

\begin{acknowledgements}
We acknowledge F.~Parhizgar and C.~Triola for useful discussions and financial support from the Swedish Research Council (Vetenskapsr\aa det, Grant No.~621-2014-3721), the Knut and Alice Wallenberg Foundation, and the European Research Council (ERC) under the European Unions Horizon 2020 research and innovation programme (ERC-2017-StG-757553).
\end{acknowledgements}

\begin{appendix}
\section{Discretization of the WNLS Hamiltonian}\label{DisW}
After taking inverse Fourier transformation in the $x$ and $z$ directions of the WNLS Hamiltonian of Eq.\,(\ref{Ham_w}), we arrive at
\bea
\bm{H}_{\text{w}} (k_y)&=&\sum\limits_r [\bm{c}_{r,k_y}^{\dagger} \left(\bm{\sigma_x} (6 - \alpha_1 - 2 \cos k_y)-\mu_{\text{w}}\bm{\sigma}_0\right)\bm{c}_{r,k_y} \nonumber\\
&&+\bm{c}_{r,k_y}^{\dagger}(-\bm{\sigma_x}-i \alpha_2 \bm{\sigma_y})\,\bm{c}_{r+\delta z,k_y}\nonumber\\
&&+ \bm{c}_{r,k_y}^{\dagger} (-\bm{\sigma_x})\,\bm{c}_{r+\delta x,k_y}+\text{H.c.}],
\label{H_w}
 \eea
where $r$$+$$\delta x$ ($\delta z$) represents the position of the nearest neighbor site along $x$ ($z$)-direction.
 $\bm{c}^{\dagger}_{r,k_y}$ ($\bm{c}_{r,k_y}$) is the creation (annihilation) operator for the electrons on site $r$ on a square lattice (in the $x$-$z$ plane) with momentum $k_y$. Note that we use $\mu_{\text{w}}$$=$$0$ throughout the paper, but qualitatively, our main results remain valid also for finite $\mu_{\text{w}}$ values. 

\section{Calculation of the anomalous Green's function}\label{anoGn}
In the Nambu basis, each part of the SC-WNLS-SC junction can be described by the Bogoliubov-deGennes (BdG) equation as
\beq
\bm{H}_{\text BdG}(k_y) \Psi(r,k_y) = E \Psi(r,k_y)
\eeq
where
\beq
\bm{H}_{BdG}(k_y)=
\begin{pmatrix}
\bm{H}_{\eta}(k_y) & \hat{\Delta} e^{i \phi} \bm{\sigma}_y \\
\hat{\Delta}^\dagger  e^{- i \phi} \bm{\sigma}_y & -\bm{H}_{\eta}^*(-k_y)
\end{pmatrix}.
\label{Hmat}
\eeq
$\bm{H}_{\eta}(k_y)$ may be the WNLS ($\bm{H}_\text{w}(k_y)$) or either of the two SC leads, $\bm{H}_{\text{sc}}(k_y)$, as well as the coupling in-between. Here $\Psi(r,k_y)$ is the four component eigenstates given in the basis $(\psi^{\dagger}_{r\uparrow}(k_y)$,$\psi^{\dagger}_{r \downarrow}(k_y)$,$\psi_{r \uparrow}(k_y)$,$\psi_{r \downarrow}(k_y))^T$. 

We model the electron part of each SC lead by a $2$D Hamiltonian
\bea
\bm{H}_{\text{sc}}(\bm{k})=-\mu_{\text{sc}}+2 t_{\text{sc}}\left(2-\cos{k_x}-\cos{k_y}\right),
\eea
where $\mu_{\text{sc}}$ and $t_{\text{sc}}$ are the chemical potential and hopping amplitude. Then, we take the inverse Fourier transformation along the $x$-direction and arrive at
\bea
\bm{H}_{\text{sc}}(k_y)&=&\sum\limits_{r_s} [\left(-\mu_{\text{sc}} +2 t_{\text{sc}}(2-\cos{k_y} )\right)\bm{b}_{r_s,k_y}^{\dagger} {\bm{b}}_{r_s,k_y}\nonumber \\
&&+ t_{\text{sc}}\,\bm{b}_{r_s,k_y}^{\dagger} \bm{b}_{r_s+\delta x,k_y} + \text{H.c.}],
\label{H_sc}
\eea
where $r_s$ denotes a lattice site index along the $x$-direction within SC lead and $\bm{b}^{\dagger}_{r,k_y}$ ($\bm{b}_{r,k_y}$) is the creation (annihilation) operator for the electrons in SC. $\Delta_{\text{s}}$ is the  SC gap parameter (zero for the WNLS, $\Delta_s$ for the SC leads) and $\phi$ is the SC phase. We use $\phi_L$ and $\phi_R$ to denote the phase factors of the left and right SC lead, respectively. We use equal phases $\Delta \phi$ ($\phi_L$=$\phi_R$$=$$0$) for all the plots of the anomalous Green's function. We show all the results for $20$ number of lattice sites of SC, which is large enough to model bulk SC leads. Throughout the work, we fix $t_{\text{sc}}$$=$$1$ and $\mu_{\text{sc}}$$=$$2$, to match the energy scale in the WNLS but with a Fermi level mismatch. Our results are not sensitive to these parameter choices nor the value of $\Delta_s$.

The coupling between each SC lead and the WNLS is given by the hopping Hamiltonian
\bea
\bm{H}_{\text{w-sc}}(k_y)=t_{\text{w-sc}}\, (\bm{c}_{1,k_y}^{\dagger} \bm{b}_{L1,k_y} +\bm{c}_{L_x,k_y}^{\dagger} \bm{b}_{R1,k_y} + \text{H.c.}), \nonumber \\
\eea
where $t_{\text{w-sc}}$ is the coupling strength between the WNLS and each SC lead. In our model, we couple the first site ($L1$) of the left lead to the first ($n_x$$=$$1$) site of the top layer of the WNLS and similarly, the first site ($R1$) of the right lead to $L_x$-th site of the top layer of the WNLS. 

We define the retarded Green's function in the Nambu basis as
\bea
\bm{G}^R(\omega,k_y)=[(\omega+i \delta) \bm{I}-\bm{H}_{\text BdG}(k_y)]^{-1}.
\eea
Following Eq.(\ref{Hmat}), ${\bm{G}}^R$ can be expressed in a $2$x$2$ block-matrix form as
\beq
\bm{G}^R(\omega,k_y)=
\begin{pmatrix}
\bm{G}^R_{ee} & \bm{G}^R_{eh} \\
\bm{G}^R_{he} & \bm{G}^R_{hh}\end{pmatrix}.
\label{Gmat}
\eeq
Each component of ${\bm{G}}^R$ is a $2N$x$2N$ ($N=L_x L_z$) dimensional matrix. The off-diagonal component, $\bm{G}^R_{eh}$, is used to calculate the amplitude of the induced pairing on the surface of the WNLS\,\cite{BalatskyMeissner}. For each site $r$, we can express the anomalous Green's function of the SC-WNLS-SC system as
\beq
F(\omega)=\sum\limits_{k_y}\bm{G}^R_{eh}(\omega,k_y)=
\begin{pmatrix}
[\bm{G}^R_{eh}]_{\uparrow \uparrow} & [\bm{G}^R_{eh}]_{\uparrow \downarrow} \\
[\bm{G}^R_{eh}]_{\downarrow \uparrow} & [\bm{G}^R_{eh}]_{\downarrow \downarrow}
\end{pmatrix}
\label{Gehmat}
\eeq
where the diagonal components correspond to the equal-spin triplet pairing and the off-diagonals provide the information for mixed-spin-triplet pairing ($[\bm{G}^R_{eh}]_{\uparrow \downarrow}$$+$$[\bm{G}^R_{eh}]_{\downarrow \uparrow}$) and spin-singlet pairing ($[\bm{G}^R_{eh}]_{\uparrow \downarrow}$$-$$[\bm{G}^R_{eh}]_{\downarrow \uparrow}$). Note that due to the periodicity along the $y$-direction, we here sum over all $k_y$ within the first BZ.

\section{spin polarization of drumhead-like surface states}\label{SS}
In this appendix we present additional results as well as an analytical calculation for the spin polarization of the surface states of the WNLS to further support the discussion of the spin density of states in the main text.
First, we show the constant energy cut using Eq.\,(\ref{Ham_w})
\begin{figure}[thb]
\centering
\includegraphics[scale=0.55]{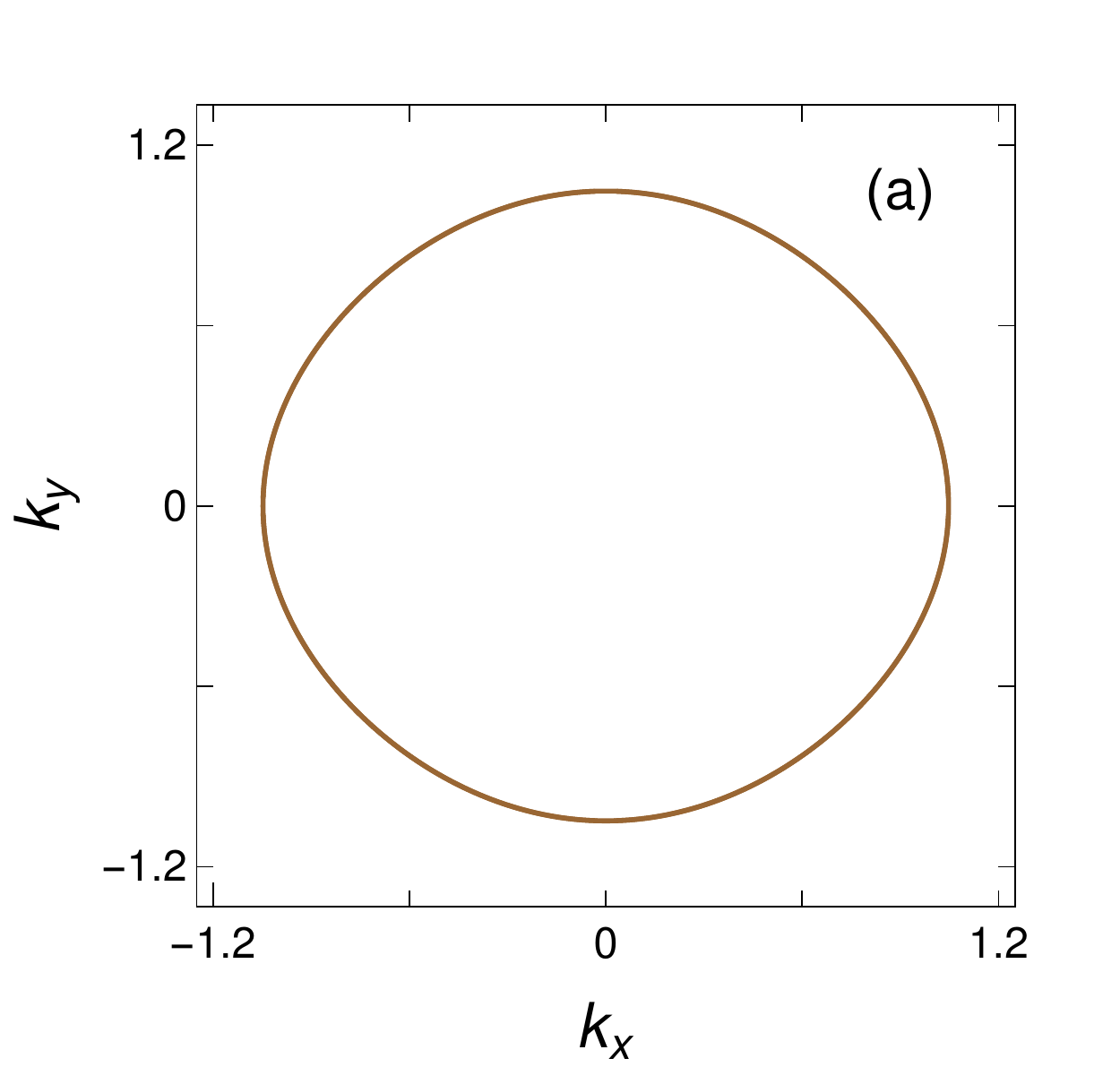}
\includegraphics[scale=0.735]{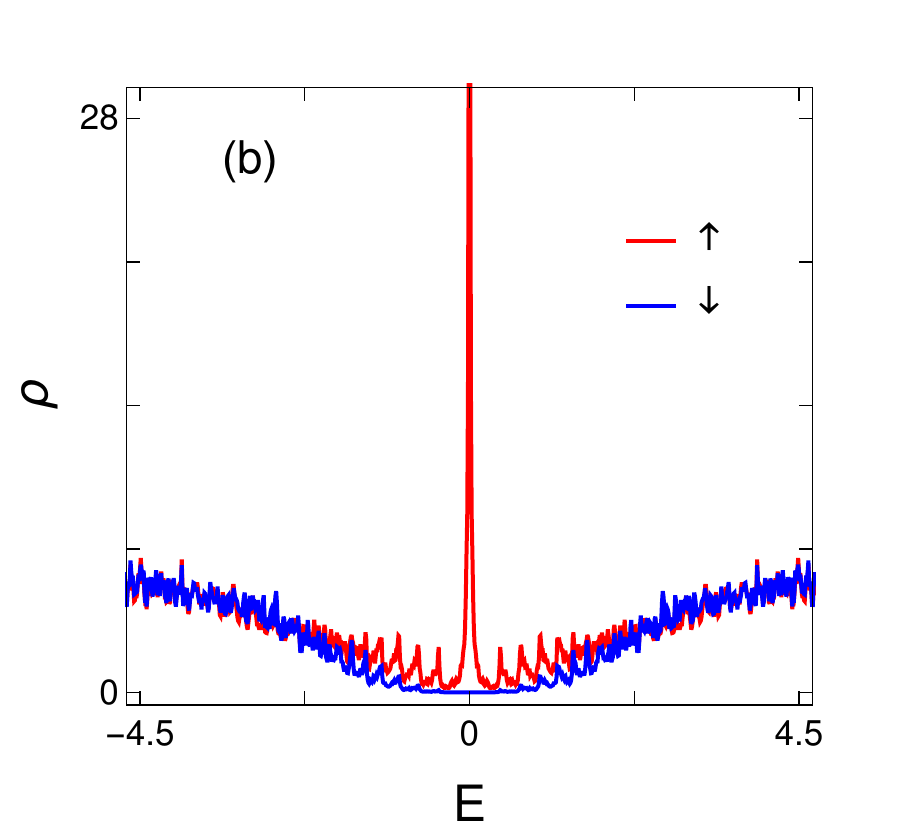}
\caption{(a) Constant energy ($E$$=$$0$) cut in the band structure in the $xy$ plane. (b) SLDOS at the middle site of the bottom layer of the WNLS as a function of energy $E$ and spin polarization. Parameters are the same as in Fig.\,\ref{model}(b).}
\label{spdos_lower}
\end{figure}
for $E$$=$$0$ in the $k_x$$-$$k_y$ plane in Fig.\,\ref{spdos_lower}(a), which confirms the Fermi nodal loop. The DSS is the projection of this nodal loop on the surface BZ, thus forming a drumhead structure.

Next, we show the behavior of the SLDOS with the variation of energy for the middle site ($L_x/2$$=$$8$) of the very bottom layer ($n_z$$=$$21$) of the WNLS in Fig.\,\ref{spdos_lower}(b). Here the SLDOS for the up spin is very 
\begin{figure*}[!thpb]
\centering
\includegraphics[scale=0.45]{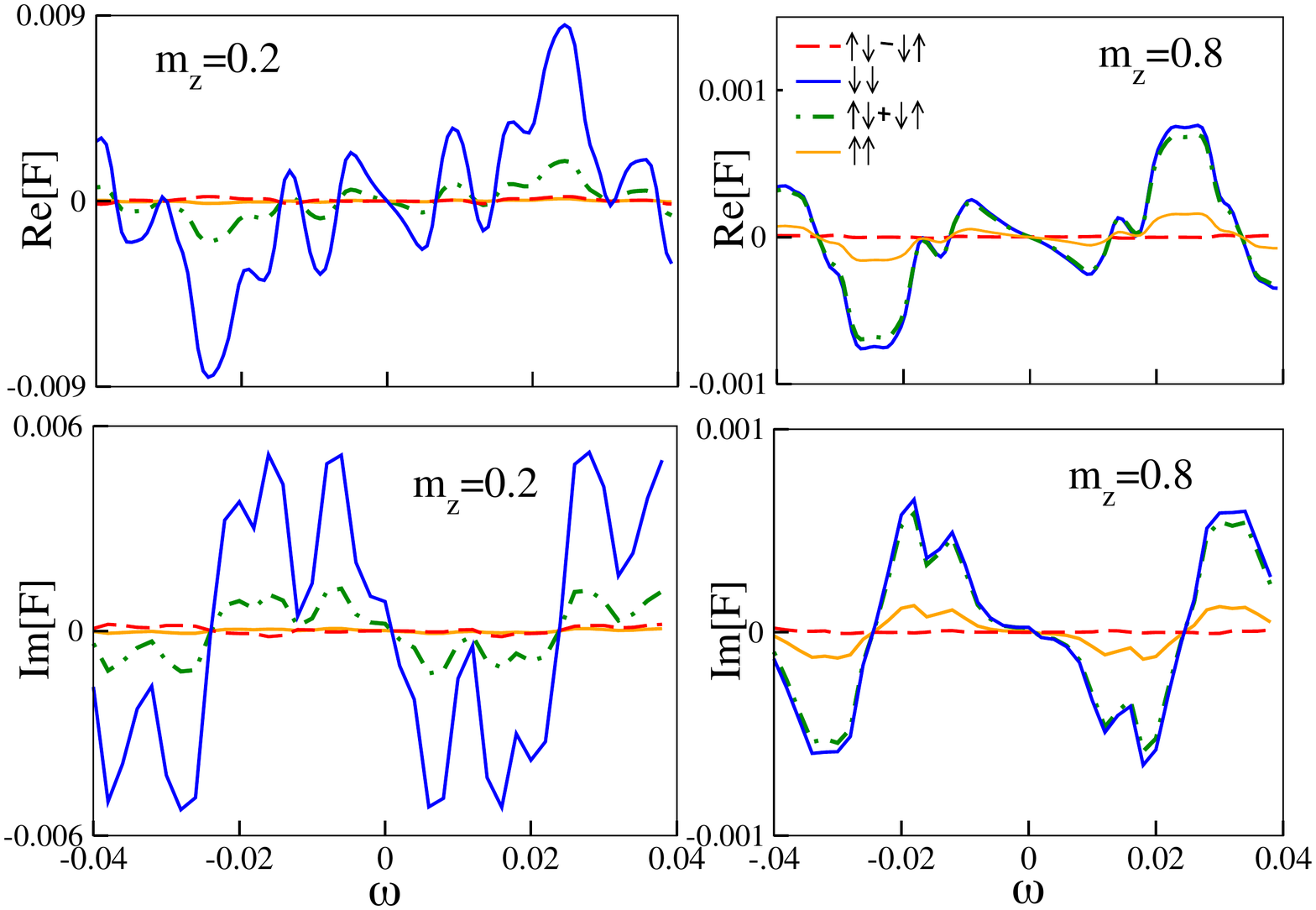}
\caption{Real (top panel) and imaginary (bottom panel) parts of $F$ as a function of frequency $\omega$ for the middle (eighth) site of the top layer ($n_z$$=$$1$) for two values of the magnetization of the FM along the $z$ axis. Other parameters  parameters are the same as in Figs.\,\ref{anoF} and \ref{Fnx}.}
\label{anoFm}
\end{figure*}
high compared to that of the down spin close to the Fermi energy. This is in complete contrast to that of the top layer as shown in the main text. Comparing Fig.\,\ref{model}(b) and \ref{spdos_lower}(b), we conclude that the surface states of the top layer are completely down spin polarized whereas those in the bottom layer are completely up spiin polarized. We also check the same for other sites of the bottom layer. The spin polarization is much higher if we move towards the middle site from the outer boundary. 

We find analytical solutions for the DSS for the Hamiltonian in Eq.(\ref{Ham_w}). Keeping both two directions $x$ and $y$ periodic, we can represent $\bm{H}_{\text{w}}(\bm{k}_{||},z)$ as an effective $1$D Hamiltonian as\,\cite{Oded},
\bea
\bm{H}_{\text{w}} (\bm{k_{||}},z)=-\bm{\sigma_x} (\partial_z^2+\alpha^{\prime}_1) -  2 i \alpha_2 \bm{\sigma_y} \partial_z 
\eea
where $\bm{k}_{||}$ (\{$\bm{k}_x, \bm{k}_y$\}), being good quantum numbers, are absorbed in $\alpha_1^{\prime}$. Further, we approximate cosine and sine functions of Eq.\,(\ref{Ham_w}) to leading order terms and replace $\bm{k}_z$ by ($-i\partial_z$). 

We set two open boundaries at $z$$=$$0$ and $z$$=$$L_z$ in order to calculate the zero-energy end state solution. The boundary conditions are expressed as
\bea
\psi_{\bm{\bm{k}_{||}}}(z)|_{z=0}=0~~ \text{and} ~~ \psi_{\bm{\bm{k}_{||}}}(z)|_{z=L_z}=0.
\label{BOC}
\eea
In the end we arrive at an equation for any zero energy states
\bea
\left[\bm{\sigma_x} (\partial_z^2+\alpha^{\prime}_1) + 2 i \alpha_2 \bm{\sigma_y} \partial_z \right]\psi_{\bm{\bm{k}_{||}}}(z)=0.
\eea
We operate $\sigma_y$ from the left side and obtain
\bea
\partial_z \psi_{\bm{k_{||}}}(z) -\frac{\bm{\sigma}_z}{2 \alpha_2} (\partial_z^{2}+\alpha_1^{\prime} )\psi_{\bm{k}_{||}}(z)=0.
\eea
Now, we look for the eigenstate of the $\bm{\sigma}_z$ operator and separate the spatial and spin parts of the wavefunction as $\psi_{\bm{k_{||}}}(z)$$=$$\phi_{\bm{k_{||}}}(z)\chi_{\nu}$ where the spin part satisfies the equation $\bm{\sigma}_y \chi_{\nu}$$=$$\nu \chi_{\nu}$, with $\nu$$=$$\pm1$.  Considering the ansatz $\phi_{\bm{k_{||}}}(z)$$\sim$$ e^{-\eta z}$, we find the secular equation as
\bea
\eta^2 \phi_{\bm{k}_{||}}(z) +2 \nu \alpha_2^{\prime} \eta \phi_{\bm{k}_{||}}(z)-2 \nu \alpha_1 \alpha_2^{\prime} \phi_{\bm{k}_{||}}(z)=0.
\eea
We have two boundaries one at $z$$=$$0$ and the other one at $L_z$ as mentioned in Eq.\,(\ref{BOC}). To find the solutions for the two surface states, we imagine two different situations. In one case, we consider it to be semi-infinite having a cut-off at the top layer at $z$$=$$0$. For infinitely large $L_z$, we can write $\phi_{\bm{k}_{||}}$$|_{L_z\rightarrow\infty}=$$0$. To satisfy this condition, the product of the two solutions $\eta_1$ and $\eta_2$ of the secular equation must be positive but we have $\eta_1 \eta_2$$=$$-$$2 \nu \alpha_{1} \alpha_2^{\prime}$. For
positive $\alpha_1$ and $\alpha_{2}^{\prime}$, we must have $\nu$$=$$-1$. This is true for all the states encircled by the projection of the bulk nodal loop on the $\bm{k}_{||}$ surface. Therefore, the DSS of the top layer of the WNLS is polarized along $-z$ direction or down spiin polarized. 

Similarly, for the bottom layer, we imagine the cut-off at $z$$=$$0$ and semi-infinite along the opposite direction \ie $\phi_{\bm{k}_{||}}$$|_{L_z\rightarrow-\infty}=$$0$. Following the similar argument, the spin polarization will be along the $+z$ direction.

\section{Frequency dependence of $F$ in the presence of FM}\label{FM-w}
In this section we provide additional data for the behavior of the superconducting pair amplitudes in the presence of a FM island in the WNLS Josephson junction, in particular, the complete frequency behavior. In Fig.\,\ref{Fnx} 
we show results for different $\bm{P}$, but choose the particular value $\omega$$=$$0.025$, as the $\downarrow\downarrow$ odd-$\omega$ pair amplitude is reasonably high at this value. We confirm the full frequency dependence we plot the real and imaginary part of $F$ in the middle site of the top layer as a function of $\omega$ for two different values of $m_z$ in Fig.\,\ref{anoFm}. From both the real and imaginary parts of $F$, it is clear that all the spin-triplet components, both equal and mixed spins, are odd in $\omega$. 

When $\bm{P}$ is set along the $z$ axis, the FM polarization partly counteracts the DSS spin polarization. This allows for both spin-singlet and mixed-spin triplet pairings to increase whereas, the amplitude of the $\downarrow\downarrow$ spin-triplet pairing is also finite due to the still strong spin polarization of the DSS. There is a competition between them and the dominant pairing is determined by the value of $m_z$. When we increase $m_z$, the amplitude of the $\downarrow\downarrow$ spin-triplet pairing decreases, while the amplitude of the  mixed-spin triplet pairing starts increasing, so that they eventually become comparable to each other. The behaviors of the real and imaginary parts of $F$ are oscillatory with the variation of $\omega$ for all the spin-triplet pairings, but clearly the result in Fig.\,\ref{Fnx} is always qualitatively valid.

\section{Calculation of Josephson current}\label{calJ}
We define a local number density operator for each site $r$ as $\bm{\hat{n}}_r(k_y)=\psi_r^{\dagger}(k_y) \psi_r(k_y)$ which provides the information for the number of particles at site $r$. It necessarily has to obey the continuity equation given by\,\cite{BlackSchaffer2008,Covaci,Kristofer},
\bea
\nabla\cdot{\bm{J}}+ e \langle \frac{\partial \bm{n}_r(k_y)}{\partial t} \rangle=0
\eea
where $\bm{J}$ is the current density vector and $e$ is the electronic charge. The time rate of change of the number density operator can be found by using the Heisenberg equation
\bea
\frac{\partial \bm{n}_r(k_y)}{\partial t} =\frac{i}{h} \left[\bm{H}(k_y),\bm{n}_r (k_y)\right],
\eea
where $\bm{H}(k_y)$ is the total Hamiltonian for the whole SC-WNLS-SC system as
\bea
\bm{H}(k_y)=\bm{H}_{\text{w}}(k_y)+\bm{H}_{\text{Lsc}}(k_y)+\bm{H}_{\text{Rsc}}(k_y)+\bm{H}_{\text{w-sc}}(k_y).\nonumber \\
\eea
 Here, $\bm{H}_{\text{Lsc}}(k_y)$ and $\bm{H}_{\text{Rsc}}(k_y)$ refer to the left and right SC lead as given in Eq.\,(\ref{H_sc}). To calculate the expectation value of the time evolution of the number density operator we consider all the occupied levels of the WNLS. The expectation value of the time evolution of the number density operator gives us two different terms proportional to the incoming [$\psi_r^{\dagger}(k_y) \psi_{r-x}(k_y)$ terms] and outgoing [$\psi_r^{\dagger}(k_y) \psi_{r+x}(k_y)$ terms] current through the nearest neighbor bonds (since all terms in the full Hamiltonian are either on-site or nearest neighbor couplings) allowing us to write the expression $\nabla \cdot \bm{J}$ as $(J_\text{out}-J_\text{in})/a$ where $a$ is the lattice constant of the unit cell. 

Note here that the sum of the currents flowing between any two neighboring site $n_x$ to $n_{x}+1$ of a particular layer $n_z$, denoted by $J_{n_z,n_x}(k_y)$,
\bea
J_{n_x}=\sum\limits_{n_z=1}^{L_z} \sum\limits_{k_y}J_{n_z,n_x} (k_y),
\label{eq:Jsum}
\eea
is constant throughout the WNLS since the current has no sinks or sources in the WNLS (the current is driven by an imposed phase difference in the SC leads only).Since $J_{n_x}$ is the same for each $n_x$ after summing over the layers, we can set the total current $J$$=$$J_{n_x}$. For the results of the current in each WNLS layer, we do not perform the summation in Eq.\,\eqref{eq:Jsum} to arrive at $J_{n_z}$. We take a summation over all $k_y$ within the first Brillioun zone for both of them.

In order to get a higher current using a smaller system size for computational reasons, we keep $\Delta_s$ a little bit higher than the value in a realistic material. To use the realistic $\Delta_s$, we have to increase the system size more to get a significant Josephson current. This assumption does not affect the qualitative behavior of our results. 

\section{Current-phase relation}\label{I-phi}
In this section we show that the maximum Josephson current is achieved at the phase difference $\Delta \phi$$=$$\phi_L$$-$$\phi_R$$=$$\pi/2$ as used in the main text.
In Fig.\,\ref{Jphase}, we display the full $J$-$\Delta\phi$ relationship. We notice that the current is maximum when the phase difference is $\Delta \phi$$=$$\pi/2$, following closely the behavior of conventional Josephson junctions. We thus use this phase difference for all the plots of the maximum current.
\begin{figure}[thb]
\centering
\includegraphics[scale=0.6]{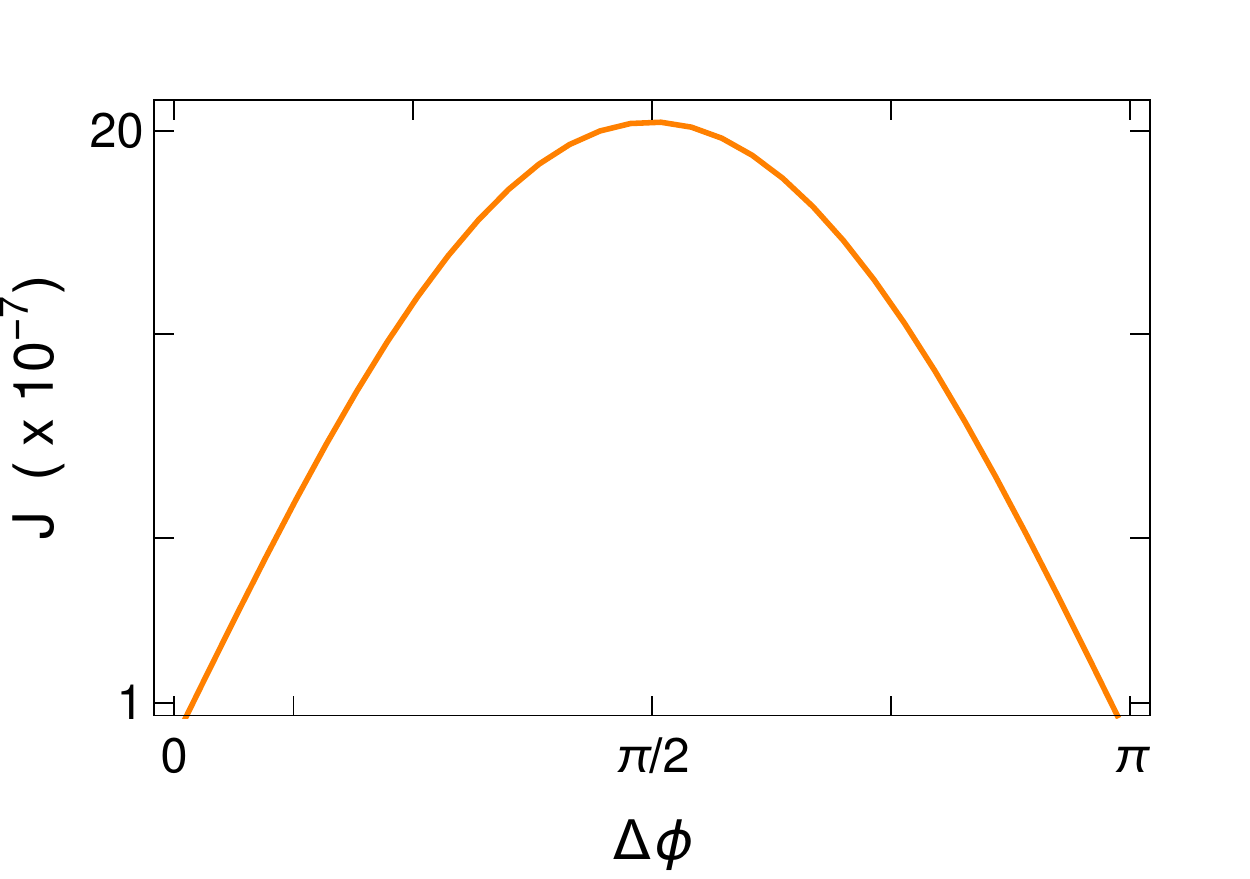}
\caption{Josephson current $J$ as a function of superconducting phase difference $\Delta\phi$ for $t_{\text{w-sc}}$$=$$0.5$ and in the absence of a FM island ($m_n=0$).}
\label{Jphase}
\end{figure}

\section{Normal metal Hamiltonian}\label{nm}
To compare the Josephson current on the surface of the WNLS with that of a normal metal Josephson junction, we model a normal metal as,
\bea
\bm{H}_{\text{NM}} (k_y)&=&\sum\limits_r[ \bm{c}_{r,k_y}^{\dagger} \left(6 - t_1 - 2 \cos k_y-\mu_{\text{NM}}\right)\bm{c}_{r,k_y}\nonumber \\
&&-t_2(\bm{c}_{r,k_y}^{\dagger}\bm{c}_{r+\delta z,k_y}+\bm{c}_{r,k_y}^{\dagger}\bm{c}_{r+\delta x,k_y}+\text{H.c.})], \nonumber \\
\label{H_nm}
 \eea
where the hopping integrals along both the $x$ and $z$ directions are $t_2$. We set $t_1$$=$$t_2$$=$$1$ and $\mu_{\text{NM}}$$=$$0$ to keep the symmetry with the WNLS Hamiltonian. This is the simplest possible normal metal state, which is also directly comparable to the WNLS Hamiltonian.

\end{appendix}

\bibliography{bibfile}

\end{document}